\documentclass[iop]{emulateapj}
\usepackage[utf8]{inputenc}
\usepackage{graphicx}
\usepackage{hyperref}
\usepackage{longtable}
\usepackage{amsmath}
\usepackage{ulem}
\usepackage{multirow}
\usepackage[dvipsnames]{xcolor}

\shorttitle{Keplerian and Eccentric Modeling}
\shortauthors{Jensen et al.}

\begin{document}

\title{Spectroastrometric Survey of Protoplanetary Disks with Inner Dust Cavities}

\author{Stanley K. Jensen Jr.}
\affil{Department of Physics and Astronomy, 118 Kinard Laboratory, Clemson University, Clemson, SC 29634-0978}

\author{Sean D. Brittain}
\affil{Department of Physics and Astronomy, 118 Kinard Laboratory, Clemson University, Clemson, SC 29634-0978}

\author{Andrea Banzatti}
\affil{Department of Physics, Texas State University, 749 N Comanche Street, San Marcos, TX 78666, USA}

\author{Joan R. Najita}
\affil{NSF's NOIRLab, 950 North Cherry Avenue, Tucson, AZ 85719}

\author{John S. Carr}
\affil{Department of Astronomy, University of Maryland, College Park, MD 20742, USA}

\author{Joshua Kern}
\affil{Department of Physics and Astronomy, 118 Kinard Laboratory, Clemson University, Clemson, SC 29634-0978}

\author{Janus Kozdon}
\affil{Department of Physics and Astronomy, 118 Kinard Laboratory, Clemson University, Clemson, SC 29634-0978}

\author{Jonathan Zrake}
\affil{Department of Physics and Astronomy, 118 Kinard Laboratory, Clemson University, Clemson, SC 29634-0978}

\author{Jeffrey Fung}
\affil{Department of Physics and Astronomy, 118 Kinard Laboratory, Clemson University, Clemson, SC 29634-0978}

\begin{abstract}
    We present high-resolution spectra and spectroastrometric (SA) measurements of fundamental ro-vibrational CO emission from nine nearby ($\lesssim$300~pc) protoplanetary disks where large inner dust cavities have been observed. The emission line profiles and SA signals are fit with a slab disk model that allows the eccentricity of the disk and intensity of the emission to vary as power laws. Six of the sources are well fit with our model, and three of these sources show asymmetric line profiles that can be fit by adopting a non-zero eccentricity. The three other sources have components in either their line profile or SA signal that are not captured by our disk model. Two of these sources (V892~Tau, CQ~Tau) have multi-epoch observations that reveal significant variability. CQ~Tau and AB~Aur have CO line profiles with centrally-peaked components that are similar to line profiles that have been interpreted as evidence of molecular gas arising from a wide-angle disk wind. Alternatively, emission from a circumplanetary disk (CPD) could also account for this component. The interpretations of these results can be clarified in the future with additional epochs that will test the variability timescale of these SA signals. We discuss the utility of using high-resolution spectroscopy for probing the dynamics of gas in the disk and the scenarios that can give rise to profiles that are not fit with a simple disk model. 
\end{abstract}

\section{Introduction}
As gas giant planets form, they interact with the disk in ways that are expected to produce observable signatures of their presence. For example, a massive planet will restrict the flow of material from the outer disk across its orbit and into the inner disk \citep[e.g.,][]{lubow1999}. This process is expected to result in disks whose inner regions reveal evidence of the depletion of dust (i.e., gapped disks, pre-transitional disks, or transition disks). Giant planets can also drive spiral arms in the disk \citep[e.g.,][]{fung2015}. Indeed, \citet{dong2018} point out that direct imaging of nearby Herbig stars indicates that $\sim20-50\%$ of intermediate-mass stars harbor a gas giant planet from beyond $\sim$30~au. 

While the depletion of dust in the inner disk and the appearance of spiral arms are enticing signposts of planet formation in disks, other effects can give rise to these features. For example, \citet{najita07} summarize three other scenarios that can in principle result in a transition disk without the presence of a planet: tidal interactions with a stellar companion, preferential grain growth in the inner disk, and photoevaporation of the disk. Similarly, stellar flybys can induce spiral patterns in disks \citep{smallwood2023}. Thus the identification of a disk whose inner region is cleared or whose disk reveals spiral structures is not sufficient to identify the presence of a planet in a disk. Unfortunately, direct imaging of gas giant planets in disks has proven difficult. To date, the reported detections of forming planets in disks remain hotly contested with the exception of PDS~70 bc \citep{keppler18, muller2018, haffert2019}. 

An alternative way to search for signposts of planet formation is the use of high-resolution spectroscopy (R$\gtrsim$25,000) of ro-vibrational molecular emission lines \citep[e.g.,][]{brittain2023}. The observation of ro-vibrational CO emission has been used extensively to study warm molecular gas in the inner disk \citep[e.g.,][]{najita2000spec}. While such emission is generally not spatially resolved, the radial distribution of the emitting gas can be inferred from the resolved spectral line profiles \citep[][]{brittain03, najita2007b, Blake2004, Salyk2011}. Warm ro-vibrational molecular emission is typically found to arise within the inner $\sim$1~au of disks around T~Tauri stars \citep{najita2003}, though it can be far more extended in the case of Herbig stars \citep{brittain03, brittain07, goto06, vanderplas15}. The availability of even higher resolution spectrometers (R$\sim$100,000), surprisingly revealed that even many resolved lines were centrally peaked \citep{bast11}. \citet{pontoppidan08} applied spectroastrometry (SA) to the study of ro-vibrational CO emission lines that provided additional information about the distribution of the gas.

SA is the measurement of the center of the point spread function (PSF) of the spectrum at each wavelength \citep[e.g.,][]{bailey98b, takami2003, brannigan2006, whelan08,brittain15}. In the case of an emission line arising from a circumstellar disk, the red- and blueshifted side of the line will be extended in opposite directions along the slit axis when the position angle of the slit is aligned with the position angle of the disk \citep[e.g.,][]{pontoppidan08, pontoppidan11}. As the slit is rotated from the semi-major to the semi-minor this offset reduces until none is observed. When the gas is not axisymmetric, the shift in the center of the PSF can be more complicated \citep[e.g.][]{brittain2013}.
\citet{brittain15} provided a review of the application of SA to the study of protoplanetary disks, including the ability to measure to small scales and usefulness in monitoring spatial aspects of spectroscopic variability.

Ro-vibrational spectra from protoplanetary disks have been analyzed with SA in several studies \citep[][]{pontoppidan08,pontoppidan11,brittain09,brittain15,brittain18,brown12,brown13,jensen2021}.
 In many cases the spectra were found to be consistent with emission arising from gas in a Keplerian orbit in an axisymmetric disk.
For high accreting systems ($\dot{M} > 10^{-7}$), \citet{pontoppidan11} found that the radial distribution of ro-vibrational CO emission was surprisingly compact, such that if line emission originated from rotating gas within the disk then the emitting gas must be sub-Keplerian. Sub-Keplerian rotation of the disk atmosphere is expected for disks that redistribute angular momentum via surface accretion flows \citep[e.g.][]{zhu2018,bai2017}. \citet{pontoppidan11} proposed that the lines observed in these systems arose from a wide-angle disk wind rotating at a sub-Keplerian orbital velocity, a possibility which has also been investigated in other studies without SA \cite[e.g.][]{bast11,banzatti2022survey}. Since the blueshift of the emission is at most a few km s$^{-1}$, the outflow velocity of the hypothesized wind cannot be large. In T~Tauri stars, outflows have long been studied using optical [OI] emisison as a tracer of low-density, hot gas \citep[e.g.][]{edwards1987, hartigan1995}. Recent studies of low-velocity components of [OI] found singly-peaked, low-blueshift lines consistent with a wind interpretation \citep[e.g.][]{fang2018,mcginnis2018,banzatti2019}. Modeling efforts have also reinforced the idea of emission blueshifted by only a few km~s$^{-1}$ relating to [OI] and molecular winds, including CO \citep[e.g.][]{wang2019,weber2020,rab2022} while the counterflow may be hidden by the circumstellar disk. While a singly-peaked, low-velocity wind can be difficult to distinguish from a Keplerian profile with spectroscopy alone, by employing SA it is possible to reveal non-Keplerian signatures which can differentiate between the two scenarios. If the CO emission from such a wind is sufficiently far from the disk surface, we would expect to detect a SA signal directed perpendicular to the disk \citep{pontoppidan11}. The degree of variability of emission lines formed in a wind is uncertain as there are limited multi-epoch observations of such sources. If the lines are variable, whether in brightness or spatial offset, it is expected that the timescale of such variability would be stochastic like that of variable accretion.

High-resolution, multi-epoch observations of CO emission revealed further surprises. CO emission from the transition disk around the Herbig star HD~100546 was found to be variable \citep{brittain09}. Subsequent observations confirmed the variability, and the application of SA enabled these authors to determine the location of the variability \citep{brittain2013, brittain14, brittain19}. These authors found that the source of the variability was a compact source of emission orbiting near the inner edge of the disk. Such a signal is consistent with emission from a circumplanetary disk (CPD).

 CPDs are expected to extend between 30\% to 100\% of the extent of the Hill sphere \citep[e.g.,][]{quillen2004, szulagyi2016-disc, Szulagyi2017}. For a 5~M$\rm_{J}$ planet 10~au from 2.5~M$\rm_{\odot}$ star, the extent of a CPD is expected to be 0.3-1.0~au and reach temperatures in excess of 500~K \citep[e.g.,][]{Szulagyi2017}. The emission lines from the CPD are expected to shift relative to the system velocity on an orbital time scale, a signature that has been observed for HD~100546 \citep[][]{brittain2013, brittain14, brittain19}. Indeed, radiation-thermochemical simulations of a CPD orbiting HD~100546 by \citet{oberg2022} reproduce the CO luminosity of the orbiting emission reported by \citet{brittain2013}. Another signpost of the presence of a forming planet results from the tidal interaction between the planet and the disk giving rise to a persistent disk eccentricity in the vicinity of the planet \citep[e.g.,][]{papaloizou2001, kley06}. The ro-vibrational OH emission lines observed from HD~100546 have an asymmetric structure that can be produced by a narrow eccentric annulus of gas \citep{liskowsky12, brittain19}.

Other effects can give rise to emission line profiles and/or SA signals that are not consistent with emission from an axisymmetric disk. For example, a vortex can also create non-axisymmetric structures in the disk that have been revealed by sub-mm imaging  \citep{vdm2021}. Three-dimensional hydrodynamical simulations indicate that a vortex extends vertically throughout the disk \citep[e.g.][]{barge16,raettig2021} or may even move away from the midplane and reside at the disk surface \citep[][]{barranco05,marcus2013}. Vortices in Keplerian disks are anticyclonic and, therefore, high-pressure regions that are elevated in temperature and/or density, depending on thermodynamics. The density and temperature fluctuations of gas have a complicated relationship with the emission arising from molecular gas at the disk surface. If the gas density is below the critical density for the vibrational levels of CO (n$\sim10^{12}$ cm$^{-3}$, e.g. \citealt{najita1996}), density fluctuations will have a strong influence on the vibrational population of the molecules. On the other hand, density enhancements may quench fluorescence \citep{Thi2013}. Temperature fluctuations in the gas can also affect the level populations, but they can also affect the scale height of the gas thereby changing the amount of UV light captured by the molecules and thus the intensity of ro-vibrational CO lines \citep[e.g.,][]{brittain07}. Regardless of whether vorticity enhances or diminishes the luminosity of CO emission, this effect will be azimuthally extended and vary on an orbital timescale. The observational consequences of these three scenarios, as well as the previously discussed wind scenario, are summarized in Table~\ref{tab:scenarios}. 

Here we present ro-vibrational CO spectra from nine disks drawn from \citet{banzatti2022survey} for which an SA signal can be measured. The majority of the emission line profiles observed among the sample can be reproduced with a slab model of a disk that treats the intensity and eccentricity of the disk as power laws. However, three sources (CQ~Tau, V892~Tau, and AB~Aur) exhibit features in their spectroscopic and SA profiles that cannot be replicated with emission from an axisymmetric rotating disk and are prime candidates for follow-up observations and more detailed modeling.

In this paper, we begin with an overview of our sample followed by a description of the observations and data reduction process. We then describe the modeling of the data and discuss the results of our modeling for the individual sources. We conclude with a summary of the results of the survey and discuss further observations that can elucidate the origin of the emission.

\section{Sample}
We selected nine intermediate-mass stars within $\sim$300~pc with evidence of gaps/large inner holes (see Table~\ref{tab:literature_params}). Eight of the targets are known to have structures such as spiral arms (HD~141569, LkH$\alpha$~330, EM*~SR~21A, CQ~Tau, V892~Tau, AB~Aur) and/or rings (HD~141569, HD~169142, EM*~SR~21A, Oph-IRS~48). Here we provide a brief overview of each target.

\subsection{HD 141569}
HD~141569 is a $\sim$2 M$_{\odot}$ \citep[][]{guzman-diaz2021survey} star with an extended disk \citep[e.g.][]{clampin03} at a distance of 110.6$\pm$0.5~pc \citep{gaiamission,gaiadr2,vioque2018dr2}. \citet{Mawet2017} find the position angle (PA) of the disk is $349\pm8\degr$ and the inclination of the disk is $53\pm6\degr$ based on scattered light $L^{\prime}$ imaging which is consistent with the PA ($356\degr$) and inclination ($51\degr$) inferred from near-infrared, scattered light observations of the outer disk acquired with NICMOS at 1.1~$\mu$m on the Hubble Space Telescope \citep{weinberger99}. We adopt these latter values to be consistent with previous modeling efforts \citep{jensen2021}.  

The inner edge of the disk was estimated to extend into $\sim$10~au based on modeling of the SED \citep[][]{malfait98} as well as various other structures at larger radii such as spirals and rings \citep[][]{clampin03,perrot16,konishi16}. This target is further noteworthy for being between the transition disk and debris disk classifications \citep[][]{wyatt15}. Previously published work, revisited in more detail below, found ro-vibrational CO emission from 12 to 60~au consistent with an axisymmetric rotating disk \citep{jensen2021}.

\subsection{HD 179218}
HD~179218 is a 3~M$_\odot$ \citep{guzman-diaz2021survey} Herbig Ae/Be star at a distance of 266.0$\pm$3.3~pc \citep{gaiamission,gaiadr2}. We adopted the disk inclination of 40$\degr$ from \citet{fedele2008} and position angle of 23$\degr$ from \citet{vanderplas15} for the analysis in this paper. Previous observation of the ro-vibrational CO emission indicates that the inner edge of the molecular component of the disk is 11~au \citep{vanderplas15,brittain18}. However, modeling of the SED is consistent with the disk extending into 0.56~au \citep{guzman-diaz2021survey}, and high contrast imaging in the $J-$band with GPI reveals no structure in the outer disk (R$\gtrsim$20~au) \citep{Rich2022}.

\subsection{HD 169142}
HD~169142 is a 1.85~M$_\odot$ \citep{gratton2019} Herbig at a distance of 114.0$\pm$0.8~pc \citep{gaiamission,gaiadr2}. We adopt a disk PA of $5\degr$ and inclination of $13\degr$ \citep{Raman2006} for the analysis in this paper.  The disk has a variety of features, including a number of rings \citep[e.g.][]{quanz13,fedele17,macias17,perez19} as well as gaps at various wavelengths, including mid-infrared and radio/millimeter \citep[e.g.][]{honda12,quanz13,osorio2014,fedele17} with a dust-depleted cavity inward of $\sim$20 au. A point-like feature in the disk has been observed over three epochs \citep{Hammond2023}. The motion of this point source is consistent with an object in a Keplerian orbit and indicates the presence of a young gas giant planet. This source of emission is 0$^{\prime\prime}$.32 from the star at a PA=44${\degr}$. 

\subsection{\texorpdfstring{LkH$\alpha$}{LkHa} 330}
LkH$\alpha$~330 is a 2-3~M$_\odot$ \citep[e.g.][]{valegard2021,pontoppidan11,andrews2018scaling} intermediate mass T~Tauri star with a (pre-)transition disk \citep{espaillat14} at a distance of 318.2$\pm$3.5~pc \citep{gaiamission,gaiadr2}. We adopt the position angle and inclination of the disk from \citet[][$28\degr$ and $49\degr$ respectively]{pinilla2022}. It is known to have a significant dust cavity from submillimeter and infrared (IR) imaging \citep{brown2008,andrews2011,isella2013,uyama2018} with a size of $\sim$50-80 au as well as indications of spiral and asymmetric structure \citep{akiyama2016, pinilla2022}. \citet{pinilla2022} find that they can account for the structures observed in the disk with a 10~M$\rm_J$ planet orbiting at 60~au with an eccentricity of 0.1.

\subsection{EM* SR 21A}
EM* SR~21A (hearafter SR~21) is a 1.64~M$_\odot$ \citep[][]{valegard2021} star at a distance of 136.4$\pm$0.6~pc \citep{gaiamission,gaiadr2}. For the analysis in this paper, we adopt the PA ($18\degr$) and inclination ($16\degr$) of the disk from \citet[][]{vandermarel2016trans-gaps}. The profile of the ro-vibrational CO emission lines indicates that the inner edge of the molecular disk is at 6.5~au \citep[][]{pontoppidan08}. Observations at millimeter wavelengths have yielded evidence of ring structure and a large cavity in dust, while there is either no gas cavity or a significant size mismatch \citep[e.g.][]{vandermarel2016trans-gaps}. Other authors have noted features such as gaps and spiral arms, including an apparent mismatch between the gas and dust extent ($\sim$7 au inner edge for the gas and $\sim$35 au clearing of large dust grains), and have been interpreted as possible signs of planet-disk interaction \citep{sallum2019,muro-arena2020}.

\subsection{Oph-IRS 48}
Oph-IRS~48 (hearafter IRS~48) is a 2 M$_\odot$ \citep{brown12} star at a distance of 134.5$\pm$2.1~pc \citep{gaiamission,gaiadr2}. For the analysis in this paper, we adopt the position angle ($95\degr$) and inclination ($42\degr$) of the disk from \cite{brown12}. Previous CO ro-vibrational observations with CRIRES showed evidence of a ring of emission co-spatial with the edge of a dust cavity at approximately 34~au \citep{brown12}. The inner edge of the disk (R$\sim$15-20~au) shows evidence of eccentricity based on dust modeling compared with 690~GHz continuum observations, which may be caused by a stellar-mass binary companion at approximately 10~au \citep{calcino2019}.

\subsection{CQ Tau}
CQ~Tau is a 1.57 M$_{\odot}$ \citep[][]{wolfer2021} star at a distance of 163.1$\pm$2.2~pc \citep[][]{parker2022}. For the analysis in this paper, we adopt the position angle and inclination of the disk from \citet[][55$\degr$ and 35$\degr$ respectively]{ubeiragabellini2019}. The disk is known to have an inner cavity of approximately 20~au as observed at millimeter wavelengths, and recent $^{12}$CO millimeter observations with ALMA \citep{wolfer2021}. These authors show evidence of non-Keplerian motion in the inner disk of the system, possibly due to a pressure profile caused by a planetary-mass companion. These authors also identified three spiral arms in their resolved images of CO emission, ranging from $\sim$10-180~au. These features may be the result of a previously hypothesized massive companion (6-9 M$_{Jup}$; \citealt{ubeiragabellini2019}) within the disk. There is also evidence to suggest a misalignment of the inner and outer disks. \citet{bohn2022} found different PAs for the semi-major axis for the inner and outer disk, being 140$^{\circ}$ and 234$^{\circ}$, respectively. Misalignment was similarly noted by \citet{safonov2022} by analyzing shadows cast within the disk.

\subsection{V892 Tau}
V892~Tau is a roughly equal mass binary system, total mass 6.0$\pm$0.2 M$_\odot$, of Herbig stars with a circumbinary disk \citep[e.g.][]{long2021} at a distance of 134.5$\pm$1.5~pc \citep{gaiamission,gaiadr2}. The orbit of the binary is eccentric (e$=$0.27) with a semi-major axis of 7.1$\pm$0.1~au and a period of 7.7$\pm$0.2~years \citep[][]{long2021}. For the analysis in this paper, we adopt a position angle of 52$\degr$ and inclination angle of 55$\degr$ \citep{long2021}. Previous observations at IR and millimeter wavelengths have noted features such as an inner cavity ($\sim$35 au in MIR) and spiral arms \citep[e.g.][]{monnier2008,pinilla2018,long2021}. 

\subsection{AB Aur}
AB~Aur is a $\sim$2.2 M$_{\odot}$ star at a distance of 155.9$\pm$0.9~pc \citep{gaiamission,gaiadr2}. For the analysis in this paper, we adopt 50$\degr$ for the position angle of the disk and 20$\degr$ for the inclination of the disk \citep{riviere-marichalar2022}. Position-velocity analysis of the rotational lines from CO isotopologues indicates that the gas has a non-Keplerian component \citep[e.g.][]{pietu2005,lin2006} associated with spiral arms at large radii ($\sim$150~au). This may be the result of interactions with a companion within the dust cavity \citep[][]{poblete2020}. The dust cavity itself has an inner edge of $\sim$70~au when observed at millimeter wavelengths \citep[e.g.][]{pietu2005}, though mid-infrared observations indicate the gas may extend as close as 0.3~au \citep[][]{difolco2009}. Recent observations have also found a point-like source that is  consistent with an embedded planet within the disk at a PA of $\sim180^{\circ}$ East of North and a separation of 0$^{\prime \prime}$.6 \citep[$\sim$93~au][]{currie2022,zhou2022}.

\section{Observations and Data} \label{sec:data}
Observations were taken between April 2017 and January 2023, inclusive, with the \textit{Infrared echelle spectrograph} (iSHELL) on the \textit{NASA Infrared Telescope Facility} (IRTF) 3.2-meter telescope at Mauna Kea, Hawaii \citep{rayner2022}. The resolution of the instrument is R$\approx$92,000 for the 0$^{\prime \prime}$.375 slit and R$\approx$60,000 for the 0$^{\prime \prime}$.75 slit \citep{banzatti2022survey}. In some sources, data for multiple PAs or epochs were taken as part of an ongoing spectroscopic survey of potentially planet-forming protoplanetary disks \citep{banzatti2022survey}. The targets were observed in the M-band (spanning 4.5 to 5.2 $\mu m$ or 1950 to 2200 cm$^{-1}$) using an ABBA nodding pattern, keeping the target in slit while nodding between A and B positions. This allowed the sky to be removed to first order by combining the images with an A-B-B+A pattern. Table~\ref{tab:obs_param} reports the observing settings for all spectra included here.

The observations were acquired to facilitate the measurement of the SA signal of the ro-vibrational CO emission lines. The center of the PSF can be measured to high precision, well below the angular resolution of the observing system, $\delta \sim \mathit{FWHM} / (2.35\cdot \mathit{SNR})$,
where $\delta$ is the accuracy of the measurement of the center of the PSF, $SNR$ is the signal-to-noise ratio of the spectrum, and $FWHM$ is the full width of the seeing profile at half of its maximum intensity \citet{whelan08}.

\cite{brannigan2006} note several sources of artifacts in spectro-astrometric signals. Some artifacts can be removed by rotating the slit 180 degrees such that instrumental effects remain constant while the astronomical source is inverted. By differencing these measurements, instrumental effects are removed while the astronomical source is averaged. This does not account for all sources of artifacts however. For example, if the PSF is comparable or narrower than the slit width, tracking errors can induce a signal. This signal is most prominent in the sharpest lines. Artifacts can also appear in telluric lines. We examined our spectra for artifacts in sources with narrow lines such as HD~169142 where the emission lines are only marginally resolved. We do not detect a signal from this source (artificial or real). Further, M-band spectra are dominated by telluric lines where artificial signals are not detected either.

The PA of the slit for each observation is reported in Table~\ref{tab:obs_param}. The majority are aligned to or near the semi-major axis (HD~141569, HD~169142, SR~21, IRS~48) with some also including observations along or near the semi-minor (HD~179218, CQ~Tau). The observations of AB~Aur were acquired with the slit aligned with the semi-major and semi-minor axes. A third observation was acquired with the PA of the slit aligned with a feature reported by \citet{boccaletti2020}. Of the remaining two disks, V892~Tau has two observations, one along the semi-major axis and one somewhat offset, while LkH$\alpha$~330 has only one observation offset from the semi-minor axis by 26$\degr$.

The data were reduced in a manner typical of IR spectroscopy (flat fielding, image stacking, spectra extraction, etc.) using a custom IDL reduction pipeline; the full reduction process is described in \citet{brittain18}. The raw frames are combined before being flat fielded. As the iSHELL instrument is an echelle spectrograph, the spectrum is split into multiple orders, and each is reduced separately. As the orders are curved, a low-order polynomial is fit to the spectrum order being reduced and then rectified. A first-pass wavelength calibration is done at this point using the Spectrum Synthesis Program telluric model (SSP; \citealt{kunde74}) as a reference. Following this, a rectangular extraction window is used to obtain A and B beam (resulting from the A and B nod positions) spectra from the corresponding nod positions which are then combined. Further refinements to the wavelength calibration are done at this point as necessary. Once this has been completed for all orders for both science and telluric standard targets, the ratio of the two stars' spectra are taken, removing telluric absorption features in the process. The spectral ranges with low flux, approximately $50\%$ or less of the continuum level as determined by the standard star, are omitted. Any final adjustments to the wavelength calibration are made at this point to ensure that the science and standard results agree with one another. Once completed, the systemic uncertainty is approximately 1~km~s$^{-1}$ for the wavelength calibration.

Measurement of the SA signal was accomplished using the same process described in \citet{jensen2021}. As in that case, a skewed Moffat function was used because it faithfully reproduced the shape of the PSF of the continuum. An initial parameter fit to the profile was found using the continuum, which it should be noted is typically thermal dust emission rather than stellar continuum except in the case of the dust-depleted HD~141569, to determine the typical shape of the PSF. We do not find evidence for extended continuum emission over the seeing. Afterward, the entire spectrum was fit, allowing a more limited range to the parameters, although with the centroid and peak values still being freely variable.

The SNR of the line profile and SA signal are enhanced by stacking multiple lines. The lines selected are the ro-vibrational v=1-0 $^{12}$C$^{16}$O lines between 2000~cm$^{-1}$ and 2200~cm$^{-1}$ (approximately 4.5 to 5.2~$\mu m$). Lines that were blended with other emission features (e.g. with ro-vibrational CO lines from other excitation bands or isotopologues) were excluded. In addition to blended lines, unexcited lines (or otherwise of extremely low SNR) and lines heavily affected by telluric absorption were also excluded. The v=1-0 $^{12}$C$^{16}$O line profiles are displayed in Figure~\ref{fig:linegallery}. It should be noted that the SA data shown in Figure~\ref{fig:linegallery} is not a spatial offset, but the pixel shift of the PSF centroid converted to the spatial scale at the target disk.

\section{Modeling}
To characterize the origin of the emission line, we attempt to fit each profile and SA signal with a simple disk model where the intensity of the emission and eccentricity of the disk are treated as power laws,

\begin{align}
   I(r)=I_0(r/a_{in})^{\alpha_I}    \\
   e(a)=e_0(a/a_{in})^{\alpha_e}
\end{align}

\noindent where $I_0$ and $e_0$ are the fiducial intensity and eccentricity of the disk, $\alpha_I$ and $\alpha_e$ are the power law indices of the intensity and eccentricity, $a_{in}$ is the semi-major axis of the innermost annulus of the disk, $a$ is the semi-major axis of an annulus, and $r$ is the distance from the star to a position within the disk. 

The disk is divided into annuli from $a_{in}$ to $a_{out}$ such that the width of each annulus corresponds to a change in velocity at perihelion of 1~km~s$^{-1}$. If the width of the annulus is greater than 1~au, then the width is set to 1~au. Similarly, each annulus is divided into segments such that the azimuthal extent of the segment corresponds to a shift in the projected velocity of 1~km~s$^{-1}$. If this results in fewer than 12 segments around the annulus, then the annulus is divided into 12 segments. The intensity, area, and projected velocity are computed for each segment where the velocity is given by,

\begin{equation}
    v(a,\theta)=\sqrt{\frac{GM_{\star}}{a(1-e(a)^2)}}(cos(\theta+\omega)+e(a)cos(\omega))sin(i)
\end{equation}

\noindent where $G$ is the gravitational constant, $M_{\star}$ is the stellar mass, $a$ is the semi-major axis of the annulus, $e$ is the eccentricity of the annulus, $\theta$ the true anomaly for all annuli, $\omega$ is the argument of periapse, and $i$ is the inclination of the disk. The profile of the emission line that arises from each segment is assumed to have a Gaussian shape with a width of 2~km~s$^{-1}$. The line from each segment of each annulus is shifted by the projected velocity and summed to create the spectral line profile. This profile is scaled to the area of the average profile that is observed. Both the synthetic spectral line profile and spectroastrometric signal are convolved with the instrument profile at the appropriate resolution (3.5~km~s$^{-1}$ for the 0$^{\prime \prime}$.375 slit, and 6~km~s$^{-1}$ for the 0$^{\prime \prime}$.75 slit).

The SA signal is calculated by measuring the distance of each segment from the star and projecting it to the plane of the sky along the slit axis. The product of the luminosity of the segment and its distance from the star is divided by the luminosity of the entire line and continuum at a given projected velocity. The distances for each velocity element of the spectral line are then summed. 

The free parameters in our model are the inner and outer boundaries of the disk emission, the fiducial eccentricity, power-law indices of the intensity and eccentricity, the argument of periapse of the disk, and the normalization of the spectral line profile. We fix the stellar mass, disk PA, disk inclination, and distance to the star based on values described in Section 2 and summarized in Table~\ref{tab:literature_params}.

In cases where there is no asymmetry in the line profile at the SNR of our data, we fix the power law index of the eccentricity $\alpha_e$ at $-3$ to approximate the radial dependence of of eccentricity found from numerical simulations for a planet/star ratio $\leq 0.003$ \cite[see][]{kley06}. The argument of periapse, $\omega$ at 0$\degr$, is set to 0$\degr$ to maximize the effect of the eccentricity on the line profile to place an upper limit on the eccentricity of the disk. An emission line from a disk will be symmetric for any eccentricity when the argument of periapse is 90$\degr$. While the eccentricity of a single disk can never be ruled out definitively, the fraction of disks that have a non-zero eccentricity can be inferred statistically from a sufficiently large sample.

\begin{figure*}[h]
    \centering
    \includegraphics[trim=0 110 0 0,clip,scale=0.8]{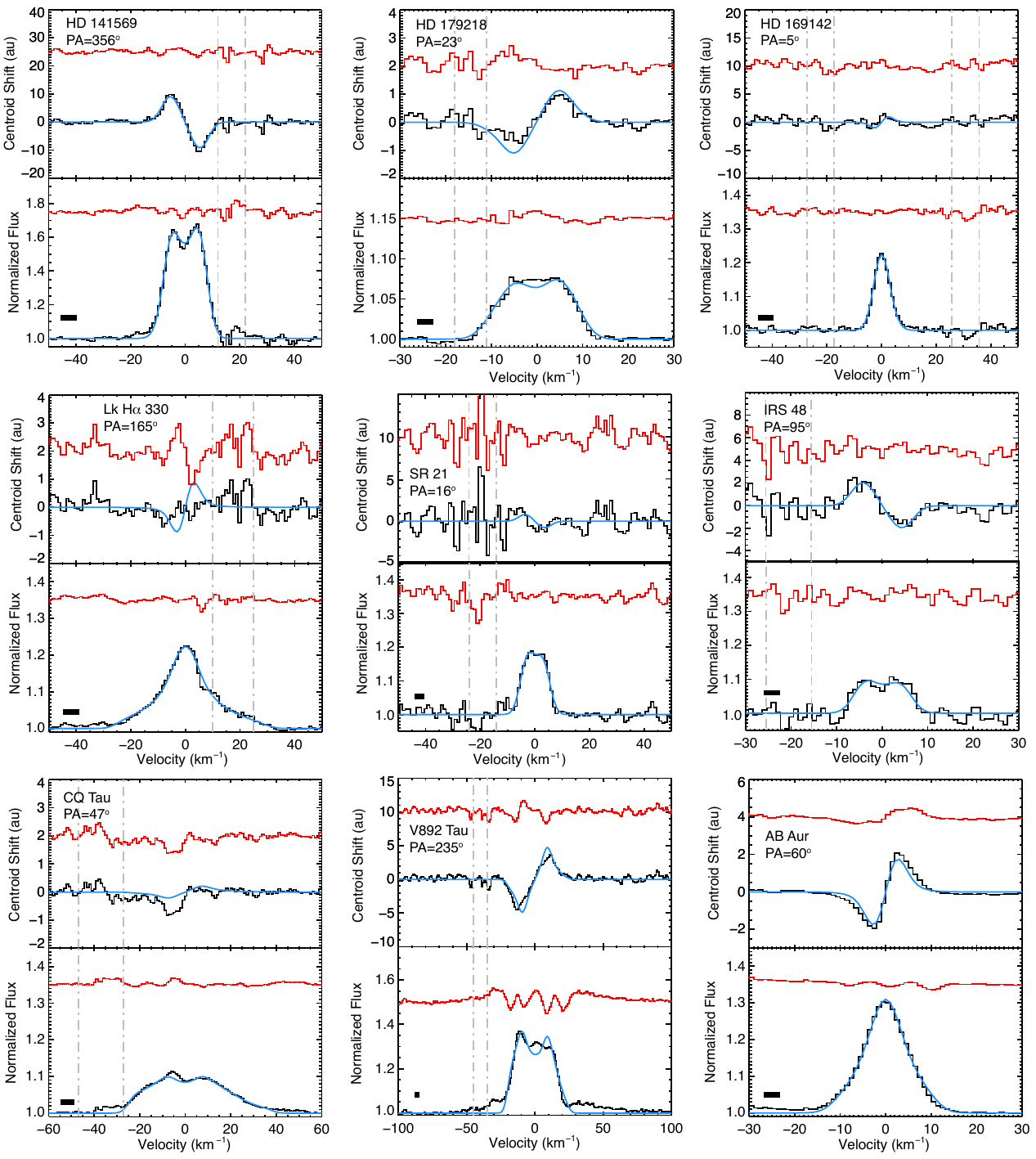}
    \caption[Gallery of Keplerian Profiles]{\label{fig:linegallery} Modeled line profiles and SA signals for the sample. The SA signal and average line profile are plotted for each target (black lines in the upper and lower panels respectively). The PA of the slit is indicated for the observation presented in the figure. The horizontal black bars indicate the instrumental resolution of the observation. The best-fit models are plotted in blue, and the residuals between the data and the fits are plotted above in red. The Doppler shift has been corrected for Earth's motion and the radial velocity of the star using the values presented in Table~\ref{tab:literature_params}. For the targets with multiple observations or epochs, those shown here are as follows: HD~169142 is averaged between Apr. 20 and Aug. 25 of 2017; CQ~Tau is from Jan. 21, 2023; V892~Tau is from Jan. 16, 2021; and AB~Aur is from Feb. 6, 2021. The velocity range affected by telluric correction of the low-J CO lines is indicated with vertical gray lines. For HD~169142, the data from two nights separated by four months were combined so two velocity ranges are indicated. The range containing telluric absorption for AB~Aur falls outside of the plotted velocity range. The SA data provide a measure of the shift of the PSF (emission line and continuum) and should not be confused with the maximum spatial offset of the emission line.}
\end{figure*}

We run a Monte Carlo simulation for each system and report the best fit and uncertainty in each parameter in Table~\ref{tab:model_fit_params}. Assuming the emission arises from a disk, the resolved sources have a unique solution. For the unresolved sources, only upper limits can be placed on certain parameters. The average spectroscopic and SA profiles along with their best-fit models are compiled in Figure~\ref{fig:linegallery}.

\section{Results} \label{sec:results}
Six of the nine objects are well fit with our disk model. CQ~Tau, AB~Aur, and V892~Tau have components that could not be fit with our model. Here we will discuss the modeling results for each target. 

\subsection{HD 141569}
The CO emission profile and SA signal are reproduced with our model assuming the emission extends from 12.5-43~au and has a relatively flat intensity profile. While the asymmetry of the line profile is quite modest, we find that we arrive at the best fit with a significant eccentricity (0.23) that declines sharply ($\propto r^{-5}$). This is significantly sharper than the decline shown in an example calculation by \citet[][$\propto r^{-3}$]{kley06} and considered in \citet{jensen2021}. The power law dependence of the eccentricity adopted in the model implies that only the first $\sim$4~au of the disk would have an eccentricity above 0.05. If the inner rim is indeed eccentric and is driven by a companion in the inner disk, the dependency of the eccentricity on the distance from the inner rim of the disk scales with the companion mass and the viscosity of the disk. Exploration of how the induced eccentricity of the disk scales with companion mass and other disk parameters will be the subject of future work. Alternatively, it could be that there is an elongated vortex at the inner rim of the disk that gives rise to the modest asymmetry of the line profile. 

One way to distinguish these scenarios is to acquire additional observations at similar resolution and SNR. Gas orbiting HD~141569 at 12.5~au has a period of $\sim$30yrs. The data presented in this paper were acquired in 2017, so in 2024 a vortex will have orbited by about 70$\degr$. The angle of periapse of the gas on an elliptical orbit driven by a companion is expected to be persistent over an orbit \citep{kley06}.

\subsection{HD 179218}
\begin{figure}[h]
    \centering
    \includegraphics[trim=50 50 20 50 ,clip,width=.5\textwidth]{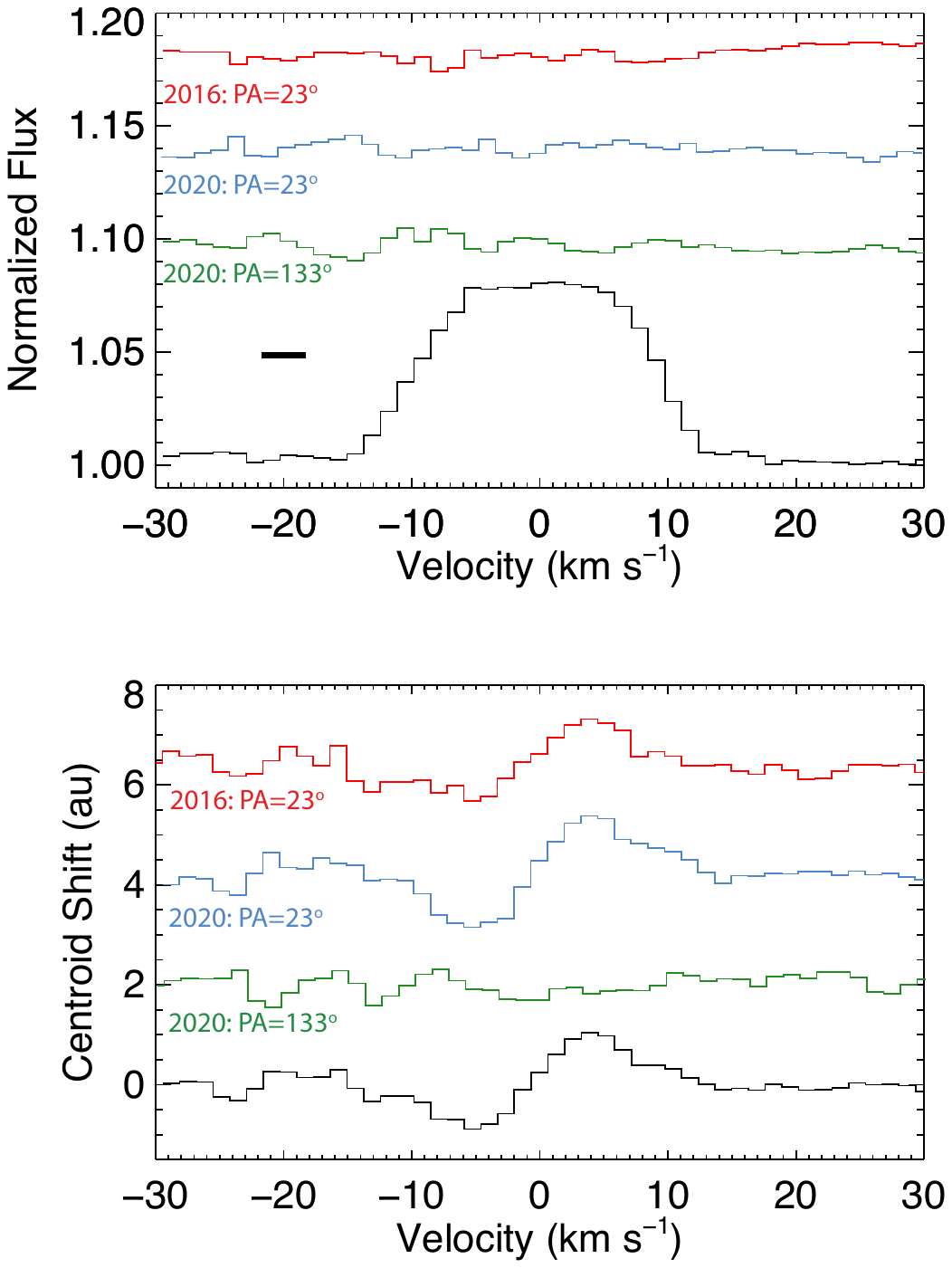}
    \caption{\label{fig:hd179} HD~179218 line profile (top panel) and SA signal (lower panel). In the upper panel, the average profile of all three observations is presented (in black). The difference between the average profile and the profile from each observation is plotted above. The horizontal black line indicates the instrumental resolution of the observation. HD~179218 was observed in 2016 \citep[][]{brittain18} with the slit aligned with the semi-major axis (in red) and in 2020 with the slit aligned with the semi-major axis (in blue) and the semi-minor axis (in green). There is no evidence of variability in the line profile - either temporally or due to changes in the orientation of the slit. In the lower panel, the SA signal for each date and PA are plotted using the same color scheme. The average of the SA signal measured with the slit aligned with the semi-major axis of the disk (in black) is also plotted. The SA signals measured in 2016 and 2020 are consistent.}
\end{figure}

The analysis of iSHELL spectra obtained in 2016 was presented in \citet{brittain18}; here we present new spectra obtained in 2020.
Spectra at two PAs, 23$\degr$ (semi-major axis) and 113$^{\circ}$ (semi-minor), were obtained for this target. The emission line and SA signal of the observation with the slit aligned with the semi-major axis were fit assuming the gas extends from 8-55~au. The eccentricity of the disk at the inner rim is 0.12 and falls off as $\propto r^{-0.8}$ - consistent with the effect of a companion orbiting interior to the inner edge of the disk \citep{kley06}. The blue wing of the line profile is extended relative to the red side. Though this fell in a velocity range that may be affected by telluric absorption, the line profile from an observation acquired in 2016 did not vary to within the noise of the spectra (Fig. \ref{fig:hd179}). Thus we conclude that the asymmetry we observe is 
 unlikely due to contamination from the telluric correction. 

\subsection{HD 169142}
The spectra of HD~169142 acquired on April 2017 and August 2017 were combined (Fig.~\ref{fig:linegallery}). The observations were acquired at the same PA, but the slit width of the observations was not the same. The effective resolution of the observations is determined by the lower resolution observations (6~km~s$^{-1}$) The profile of the emission line is only marginally resolved and no SA signal was detected. The line profile was fit assuming the emission extends from 2.7~au to 80~au. The SA signal from the modeled emission falls below the SNR of the measurement (which limits the outer extent of the centrally peaked line). Because the line is so narrow and the emission extended, our upper limit on the eccentricity is 0.25.  

A companion has been reported for this target with a separation of $\sim$0$^{\prime\prime}$.3 and a position angle of $\sim$45$\degr$ \citep{Hammond2023}. The position angle of our slit on the sky was 5$\degr$. If the candidate object has a disk, some of the emission from the disk could be captured by our slit. However, the orientation of the slit was not optimized for detection from this source. Here we assume that the continuum is coming from warm dust near the star and that any continuum emission from the CSD is insignificant. We also assume that 50\% of the CO emission comes from the CPD while the other 50\% comes form an axisymmetric CSD. The spatial offset of the candidate along the slit for this is $\sim$230~mas when taking into account the 40$\degr$ difference in the reported CPD PA and our observed PA. With the line-to-continuum ratio of $\sim$0.2 the expected SA offset would be $230\times0.5\times0.2/(1+0.2)=19$~mas or $\sim$2~au, which is below our detection threshold. \citep[see, e.g.,][for the effect of line-to-continuum ratio in the context of SA]{takami2003}. As such while we do not detect evidence of a CPD here, we cannot rule it out due to the unoptimized nature of our observations. Deeper observation of HD~169142 with the slit aligned with the position angle of the candidate CPD would in principle enable its detection from the SA signal. While a sufficiently massive companion should drive an eccentricity in the disk in the vicinity of its orbit, the large separation and face-on orientation of the disk would render any asymmetry induced in the line profile from this region of the disk difficult to detect with iSHELL in either the spectral profile because the line is so narrow ($\sim$7~km~s$^{-1}$) while the SA signal would be comparatively easy to detect with sufficient fidelity and proper alignment.

\subsection{\texorpdfstring{LkH$\alpha$}{LkHa} 330}
\begin{figure}[h]
    \centering
    \includegraphics[trim=40 70 40 70 ,clip,width=0.5\textwidth]{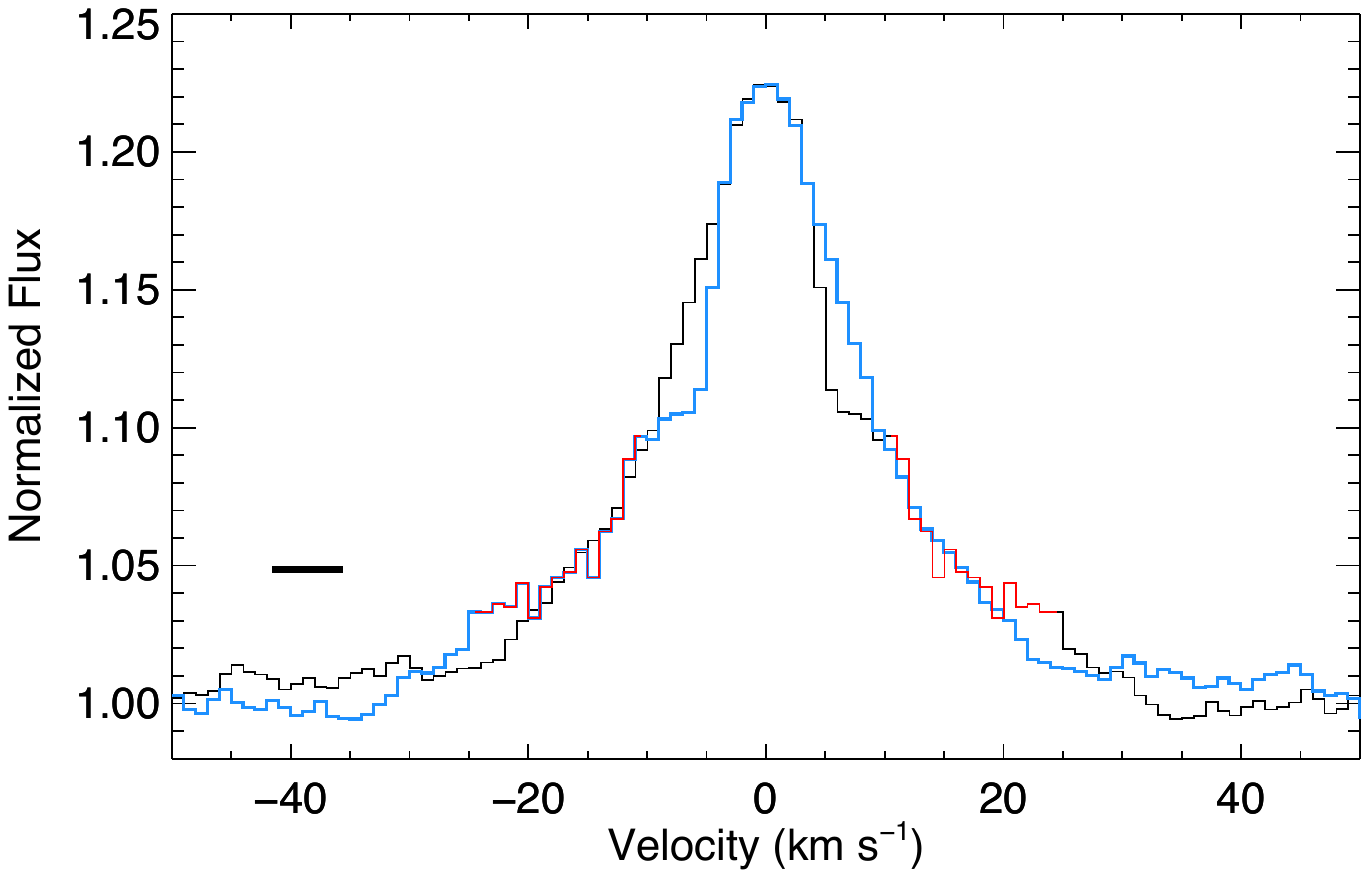}
    \caption{\label{fig:lkha330_asym} LkH$\alpha$~330 profile (black) and its reflection (blue) showing asymmetric structure. The velocity ranges covering the low-J v=1-0 telluric lines are indicated in red. The horizontal black line indicates the instrumental resolution of the observation. The core of the and the wings reveal a significant asymmetry. The HWZI of the blue side of the line occurs near $-20$km ~s$^{-1}$ while the HWZI of the red side of the line falls near $+30$km~s$^{-1}$. The core of the line shows an asymmetry in the opposite direction.}
\end{figure}

The spectroscopic profile of LkH$\alpha$~330 is broad and asymmetric (Fig.~\ref{fig:lkha330_asym}). The red wing of the line extends further than the blue wing, however, the core of the line is extended further to the blue than to the red. 
The line profile from \citet[][]{pontoppidan11} also showed asymmetry, but was significantly narrower with a sharper peak \citep[see comparison of profiles in Figure 8 in][]{banzatti2022survey}. The SA signal detected by \citet[][]{pontoppidan11} is not observed due to the lower SNR of our observation. The line profile and upper limit on the SA signal were fit assuming the emission extended from 0.46-120~au. The gas near the inner edge of the disk in our model has a modest eccentricity (0.08) though it is consistent with zero at $3\sigma$. Interestingly, we were unable to find a fit that captured the asymmetry of the wings and the core at the same time (see the residual in Fig.~\ref{fig:lkha330_asym} at +7~km~s$^{-1}$). One way to capture this is by allowing the argument of periapse to shift phases in the disk (see for example, \citealt{kozdon2023}). Such a scenario is expected when a companion opens a gap in the disk as the argument of periapse will precess at different rates at the inner and outer boundaries of the gap. Whether this is consistent with the presence of the 10~M$\rm_J$ planet at $\sim$60~au hypothesized by \citet{pinilla2022} will be explored in future work.

\subsection{SR 21}
SR~21 includes a central narrow absorption in the low-J lines assumed to be from foreground gas, so these lines are excluded from the average spectral and SA profiles presented in Fig.~\ref{fig:linegallery}. The best fit model indicates that the emission arises from a narrow annulus ($\Delta$R=1.7~au). The blue side of the line is somewhat brighter (consistent with \citet[][]{pontoppidan08}), though at the SNR of our spectrum this is only marginally significant. We can rule out an eccentricity greater than 0.13 with our data assuming $\omega$=0$\degr$. The SA signal from our model has a maximum offset of
1~au. The PA of our observations is offset by 7$\degr$ relative to the PA presented by \citet[][]{pontoppidan08} that found a signal that extended to 1.3~au (updated for the distance determined by Gaia). 

\subsection{IRS 48}
IRS~48 includes a central narrow absorption in the low-J lines, so these lines are excluded from the average spectral and SA profiles presented in Fig \ref{fig:linegallery}. The modeling results for IRS~48, shown in Fig.~\ref{fig:linegallery}, are consistent with an axisymmetric disk with an emitting region extending from 20 to 35 au -- consistent with the gas ring noted by \citet{brown12} based on modeling of CRIRES ro-vibrational CO data, including spectroastrometry. Additionally we do not find evidence of variability compared to this previous work.
The signal-to-noise of the data presented here is insufficient to detect the existence of an eccentricity in the inner rim, with an upper limit of e$<$0.17 assuming $\omega$=0$\degr$. The SA signal is well reproduced by our model. 

\subsection{CQ Tau}
\begin{figure}[h]
    \centering
    \includegraphics[trim=85 55 70 50,clip,width=.45\textwidth]{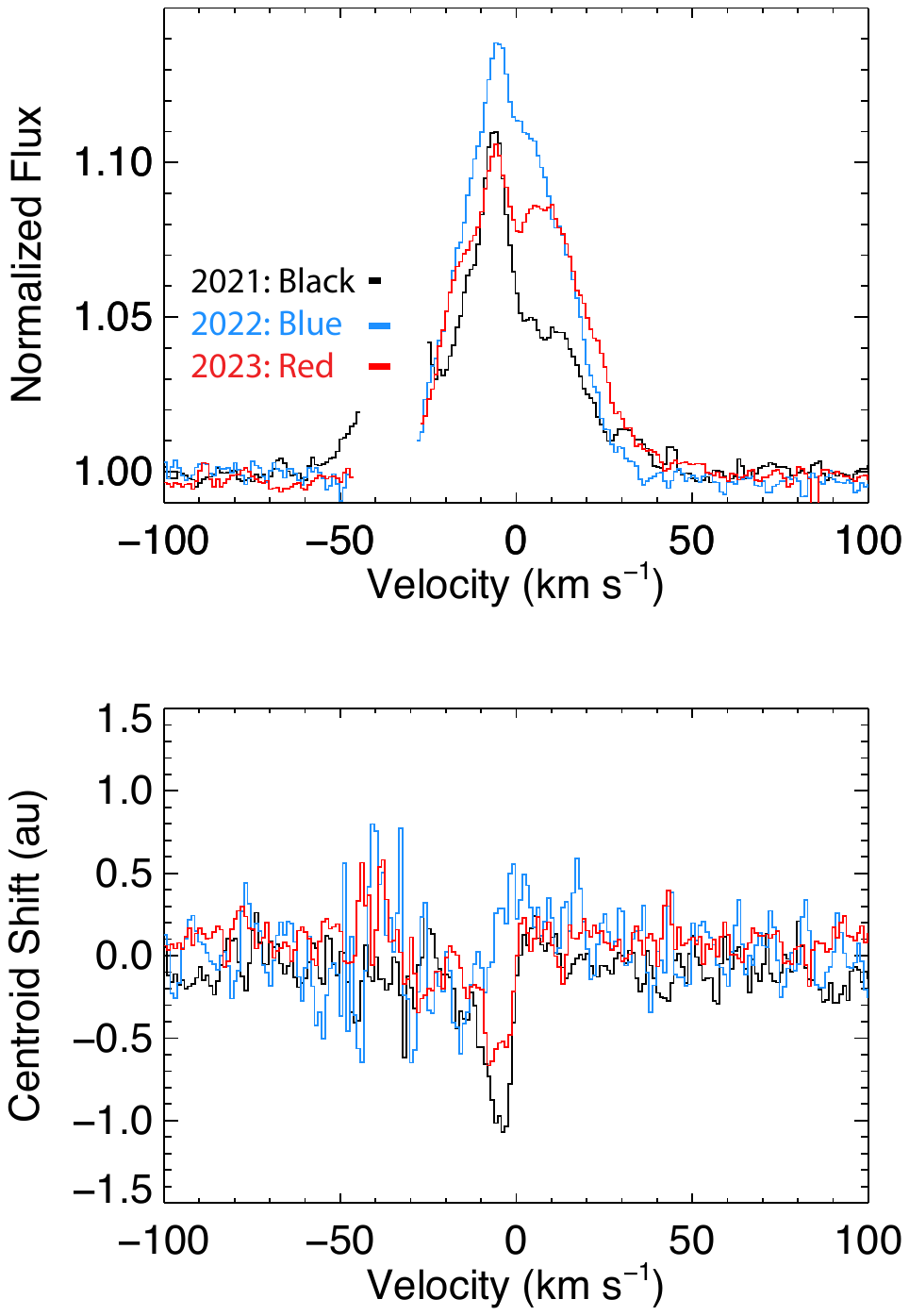}
    \caption{\label{fig:cqtau_epoch} Comparison of the spectroscopic and SA profiles of CQ~Tau observed in 2021 (black), 2022 (blue), and 2023 (red). The horizontal bars are scaled to the instrumental resolutional of each observation. A narrow emission component persists in all three epochs, though its line/continuum ratio varies. The centroid of the PSF of the narrow component is extended when the slit 47$\degr$ E of N (2021 and 2023) and disappears when the slit PA is 137$\degr$ E of N. }
\end{figure}

The data presented here were obtained in January 2021 with a slit PA of 47$\degr$ (near the semi-major axis, PA=55$\degr$; \citealt{ubeiragabellini2019}) and a slit width of 0$^{\prime \prime}$.375, in January 2022 with a PA of 137$^\circ$ (near the semi-minor axis) and a slit width of 0$^{\prime \prime}$.75, and January 2023 with a PA of 47$\degr$ and a slit width of 0$^{\prime \prime}$.75 (Table~\ref{tab:obs_param}). The profile of the emission lines and the SA signal from each epoch are complex and variable (Fig. \ref{fig:cqtau_epoch}). In all three epochs, there is a narrow component that is blueshifted 5~km~s$^{-1}$ relative to the star. In 2021 and 2023 when the slit PA was 47$^\circ$, the centroid of the PSF of the narrow component was shifted along the slit in the southwest direction. The broad component of the line profile of CQ~Tau in 2023 can be fit if we mask the narrow emission component (Fig. \ref{fig:linegallery}). For this epoch, we find that the profile can be fit with emission that extends from 0.5-9.8~au assuming the inner rim is highly eccentric (e=0.27) and that the eccentricity drops rapidly ($\rm \propto r^{-5}$). The SA signal produced by the model is consistent with the upper limit across 
the broad component of the profile. A wind may be able to qualitatively explain the nature of the narrow component and the corresponding SA signal \citep{pontoppidan11}. However, rather than the expected maximal SA offset for the CO emission when observed along the semi-minor axis of the disk for a wind, we found the offset of the SA signal was minimal along that direction while it was much more pronounced along the semi-major axis. Alternatively, taking into account evidence of misalignment of the inner and outer disks by approximately 90$\degr$ \citep{bohn2022,safonov2022}, this is consistent with a wind assuming it is being launched from the inner disk. At present it is not clear what the timescale of variability of the emission is (orbital, stochastic, or otherwise), but further epochs may clarify this.

\subsection{V892 Tau}
\begin{figure}[h]
    \centering
    \includegraphics[trim=70 55 45 50,clip,width=0.45\textwidth]{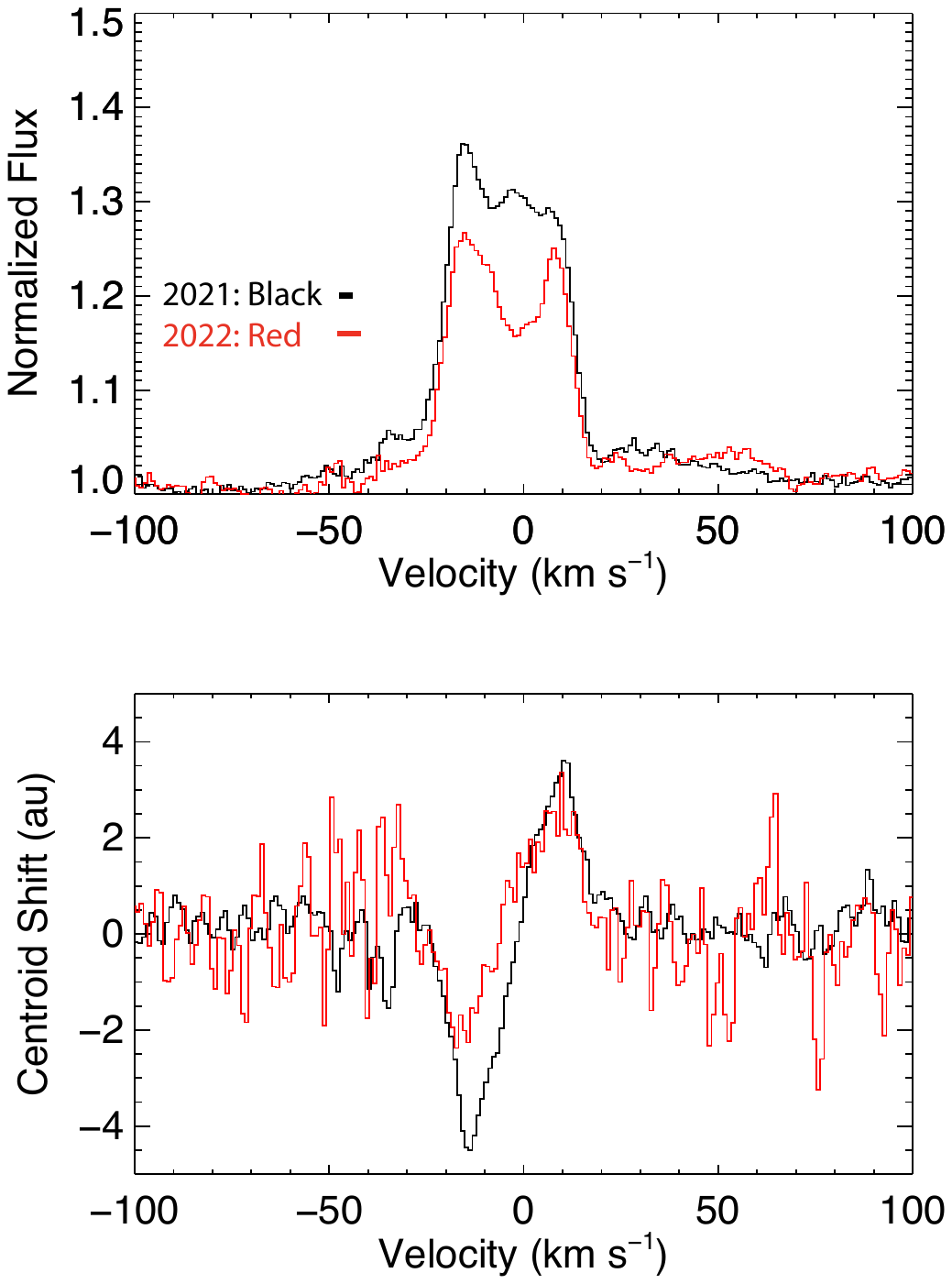}
    \caption{\label{fig:v892tau_epoch} Comparison of 2021 (black) and 2022 (red) average spectroscopic and SA profiles of V892~Tau. The horizontal bars indicate the instrumental resolution of each observation. The PA is offset by 35$^{\circ}$ between epochs. While the line profile is asymmertic in both epochs, the overall spectroscopic profile changed to exhibit a double-peaked structure, though still with an excess on the blue side. The SA signal is extended on the red and blue side of the line consistent with emission arising from an extended disk.}
\end{figure}

The CO line profile was previously shown to be highly variable on 1yr timescales \citep{banzatti2022survey}. We find that the SA signal varies with profile of the line as well (Fig. ~\ref{fig:v892tau_epoch}). The data acquired in 2021  shows an asymmetry with a single peak on the blue side of the line. In 2022 the line showed an asymmetric and double-peaked line profile. We could not find a parameter combination that could fit either profile well. A model for a chosen set of parameters for the 2021 epoch is displayed for comparison purposes (Fig.~\ref{fig:linegallery}).  The SA signal from the model is similar in extent to the measurement, but it is not a good fit. If the wings of the line arises from the circumbinary disk, the gas must extend inward to $\sim$7~au - significantly interior to the continuum inferred by MIR interferometry \citep{monnier2008}.

\subsection{AB Aur}
\begin{figure*}[h]
    \centering
    \includegraphics[trim=0 270 40 0,clip,scale=0.65]{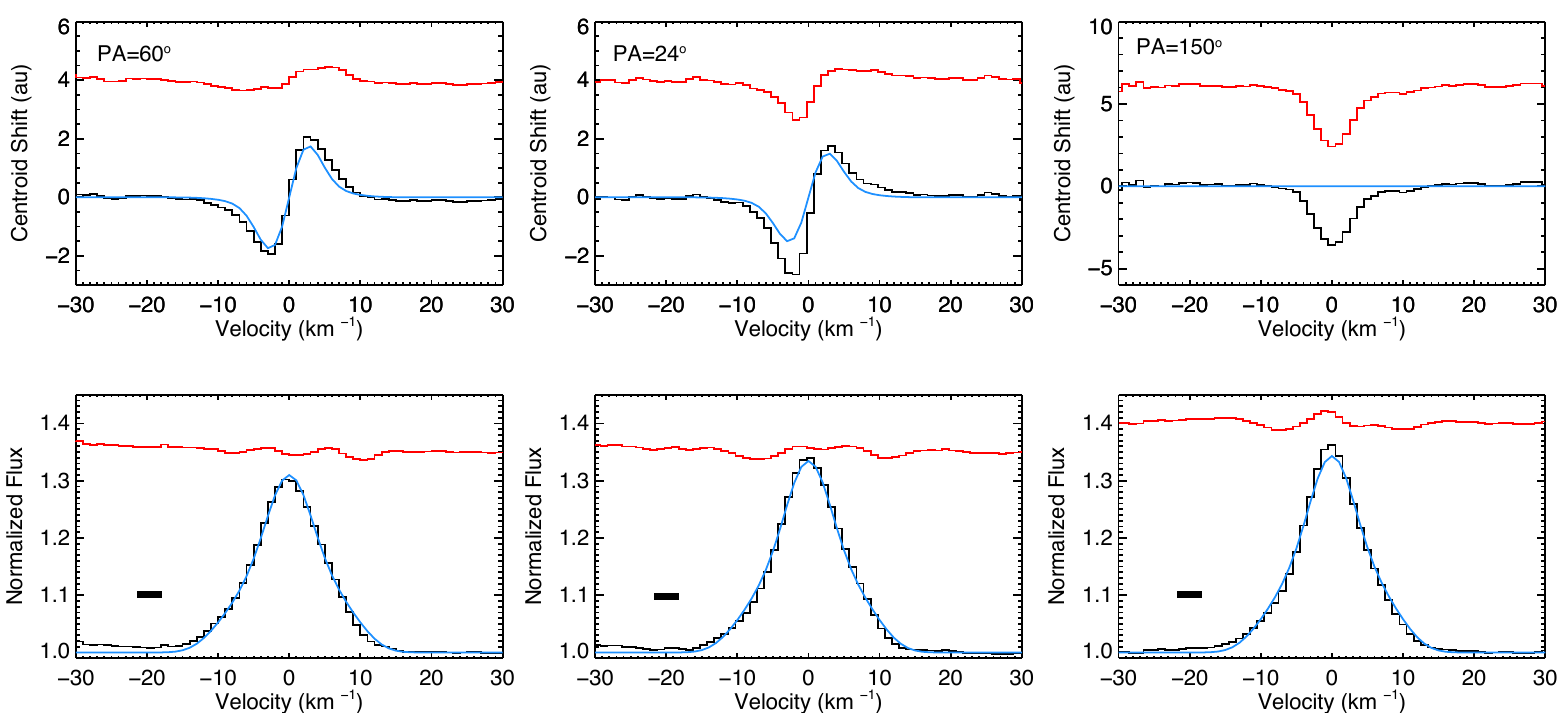}
    \caption{\label{fig:ABAur} Comparison of AB Aur observed at three different slit PAs. The data are plotted with black lines, the model is blue and the residual is plotted above in red. The horizontal lines in the lower panels indicate the instrumental resolution of the observation. The model used to fit the profile and SA signal for the slit aligned with the semi-major axis of the disk (PA=60$\degr$) was applied for two other PAs. While the profile could be fit reasonably well, the SA signal diverged significantly from the model. One way to account for this is for a non-axisymmetric source of emission on the southern part of the disk. }
\end{figure*}

The observations for this target were taken at three separate PAs: 
aligned with the semi-major axis (60$\degr$; Feb. 6, 2021), semi-minor axis (150$^{\circ}$; Feb. 4, 2021), and the position of a possible feature noted by \citet[][24$\degr$; Feb. 5, 2021]{boccaletti2020}. The line profile and SA signal acquired with the slit aligned with the semi-major axis can be reproduced by our model for emission that extends from 1-106~au. The line is quite symmetric and the upper limit we place on the eccentricity (assuming $\omega$=0) is e$<$0.05. For an axisymmetric disk, the SA should vanish when the slit is aligned with the semi-minor axis (PA=150$\degr$) and diminish when the slit is in an intermediate orientation (PA=24$\degr$). This is not observed. The model is a poor fit to the SA signal when observed with the slit aligned with the semi-minor axis (Fig. \ref{fig:ABAur}). 

During the observations (Feb 4 - Feb 6, 2021), the seeing improved dramatically (from 1$^{\prime\prime}$.76 on the first night to 0$^{\prime\prime}$.62 by the third night). If the behavior of the SA signal is due to emission from a CPD around the object observed by \citet{currie2022} at a PA$\sim$180$\degr$, it is possible that more of the emission from this source was captured in the slit on the first night than on subsequent nights explaining why the effect was larger when the slit was misaligned with the candidate planet by 30$\degr$ on night 1 than when the slit was misaligned with the candidate by 24$\degr$ on night 2. Additionally it is possible that a vortex in the disk can lead to an asymmetry in the surface CO emission. Alternatively, a disk wind can lead to a signal qualitatively similar to what is observed here \citep{pontoppidan11}. The emission line is centrally peaked, and the SA signal is extended along the semi-minor axis of the disk as expected in the wind case. Previous works have found evidence supporting a disk wind from H$\alpha$ observations and modeling of the NIR excess \citep{bans&konigl2012,perraut2016}. Further observations are planned to investigate these possibilities.

\section{Discussion}
Of the sample, six of the nine disks (HD~141569, HD~179218, HD~169142, LkH$\alpha$~330, SR~21, and IRS~48) have spectral profiles and SA signals that can be reproduced by assuming the gas arises from a circumstellar disk. Of these six, the two profiles that are centrally peaked are also nearly face-on (HD~169142 with an inclination of 13$\degr$ and LkH$\alpha$~330 with an inclination of 28$\degr$). The outer disk of SR~21 is also nearly face on (i=18$\degr$) indicating that the emission arises from a narrow annulus. In the cases of HD~169142, SR~21, IRS~48, and AB~Aur only upper limits to the inner edge eccentricity, e$_0$, were determined using an argument of periapse, $\omega$, of 0 as the most favorable orientation with a power law index, $\alpha_e$, of -3, as in \citet{kley06}. For these six cases, we cannot at present rule out alternative explanations to a gap-opening planet or CPD such as a (sub-)stellar companion, vortex, or outflow. Improved or better optimized observations, additional epochs, or more detailed modeling may better distinguish between the various possible explanations for the cavities/gaps present in this sample.

For all but two of our sources, the CO emission must extend to very large distances from the star ($\sim$50-100~au). This is quite different from T~Tauri stars where the emission typically arises from the inner few au \citep{najita2003}. The excitation of molecular gas by the much stronger UV continuum of these earlier type stars might account for the difference in the radial extent of the emission. 

The radial extent of the source of the ro-vibrational CO emission is not always aligned with the radial extent of the dust or gas as measured in the (sub)mm. One such case is HD~169142, where ALMA observations show a dust cavity inward of approximately 20~au \citep{Hammond2023} while the ro-vibrational CO emission extends inward to 2.7~au. Such discrepancies have been noted previously \citep[e.g.][]{banzatti2018}. While an orbiting companion can open a gap in the disk resulting in a pressure bump that can trap mm-sized grains, gas may accrete inward. The transition strength of ro-vibrational lines is about six-order of magnitude larger than for rotational lines in the ground vibrational state, thus the observations of the former at NIR wavelengths for CO are more sensitive to warm gas than the latter at submm wavelengths.

At the SNR of our observations, four of the observed line profiles are symmetric and upper limits on the eccentricity have been set assuming that the argument of the periapse is 0$\degr$. Of the five sources with asymmetric line profiles, the line profile and SA signal of three of the sources can be fit with a model of an eccentric circumstellar disk. 

The line profile and SA signals of AB~Aur, CQ~Tau, and V892~Tau cannot be fit with a simple disk model. Some other non-axisymmetric component is necessary beyond that caused by eccentricity. In principle, emission from a CPD could account for the SA signal of AB~Aur similarly to what observed in HD~100546 \citep[e.g.,][]{brittain19}. Alternatively, a vortex at a large distance or a disk wind  could account for the line profile and observed SA signal (Table~\ref{tab:scenarios}). These three scenarios can be distinguished with further observations and is the subject of ongoing work.

As noted above, CQ~Tau has been hypothesized to harbor a 6-9 M$_{Jup}$ planet 20~au from the star based on structures in the disk \citep{ubeiragabellini2019}. If so, it is unlikely to affect the broad component of the molecular emission profile as the emission only extends to $\sim$10~au. Given the large eccentricity necessary to account for the observed line asymmetry, it is possible that CQ~Tau harbors a close-in (sub-)stellar mass companion. The non-variable narrow component whose emission is extended along the slit when it is aligned with the semi-major remains a mystery. The line does not appear to shift more than 1~km~s$^{-1}$ over two years. However, if it arises from a CPD orbiting the star at 20~au, the orbital phase of the companion would only shift 10$\degr$ resulting in a projected Doppler shift below our detection threshold. An alternative explanation for the narrow component and SA signal is a low-velocity disk wind. However, the extension of the SA signal is a maximum when the slit is oriented near the semi-major axis, and it is not detected when the slit is aligned with the semi-minor axis in conflict with what is expected for low-velocity disk winds \citep{pontoppidan11}. The broad component of the spectral profile is the more variable of the two, but the narrow SA signal also appears to be variable as well. The complexity of the profiles over our three epochs makes it difficult to infer the distribution of the gas giving rise to these phenomena. Measuring the line profile at multiple PAs in a single night as well as long term monitoring may  clarify the origin of the SA signal. 

V892~Tau is an equal mass binary, and the dust component of the circumbinary disk extends into $\sim$17~au  \citep{monnier2008}.  The wings of the CO profile indicate that the gas extends into about 7~au - substantially inward of the inner edge of the dust component of the circumbinary disk. The orbital period of the gas at 7~au is 8 years, so the observations presented in this paper, spaced by 1 year, correspond to an orbital shift in the disk of 45$\degr$. With just three epochs, it is impossible to infer whether the CO lines vary stochastically on an orbital timescale. Observations are planned to monitor this object over six months and consecutive nights to better constrain the timescale of the variability.  

\section{Conclusions}
SA combined with high resolution spectroscopy of warm molecules in disks provides a powerful probe of the conditions of gas in the disk. We find that warm molecular emission can extend 10's of au from Herbig stars. Our sample was selected on the basis of various signposts of a companion (e.g., spiral structure or gaps). Four of our nine sources have asymmetric line profiles that can be fit assuming that the emission arises from gas on an eccentric orbit indicative of companions near the inner edge of the disk. HD~169142 is nearly face-on so the emission is only marginally resolved at our resolution of 3.5~km~s$^{-1}$. A candidate planet was identified after our observations were acquired, and our slit orientation was not favorably aligned to detect emission from this companion. While AB~Aur has a symmetric profile, the SA signal indicates non-axisymmetric structure in the outer disk. The origin of this structure will be tested with further observations. The SNR of IRS~48 and SR~21 are insufficient to place strong constraints on the eccentricity of the disk. Deeper observations at similar resolutions with instruments such as CRIRES+ on the VLT will provide stronger constraints on any line asymmetry. CQ~Tau reveals a complex line profile. The broad component varies significantly and does not reveal an SA signal. The narrow component is less variable, and it is extended. The origin of these components remain unclear. V892~Tau is an equal mass binary for which the CO line profile and SA signal is highly variable. Monitoring of this object will clarify the cadence of the variability and thus possibly the origin of the emission. LkH$\alpha$~330 has a complex profile that indicates that the argument of periapse of the eccentricity flips. The profile is also variable. A higher fidelity line profile and modeling that allows for the argument of periapse to flip may indicate the presence of a companion embedded in the disk.

CQ~Tau and AB~Aur individually represent interesting signals warranting continued observation in the context of potentially planet-harboring disks. However, they also are worthwhile testbeds for the use of SA to distinguish between such an origin as well as other scenarios such as disk winds. Winds in particular can have coincidentally similar spectral and SA signals in single epochs, but can vary drastically at other times and on different variability timescales (orbital timescales for planets or CPDs, stochastic timescales for winds). Continued investigation and monitoring of these and similar sources may result in better methods for determining the origin of the emission.

The sample presented here suggests that disks with an eccentric component are common among Herbigs with evidence of ongoing planet formation. Confirmation of our interpretation will provide a way to gain statistical information about giant planets in disks in large sample of stars at distances too far to detect such objects with direct imaging. New and upcoming instruments (e.g. METIS, \citealt{brandl2008metis}) on 30m class telescopes will be able to provide improved observations of this kind by providing a sensitivity increase by an order of magnitude.

\section*{Acknowledgements}
This publication makes use of data products from the Wide-field Infrared Survey Explorer, which is a joint project of the University of California, Los Angeles, and the Jet Propulsion Laboratory/California Institute of Technology, funded by the National Aeronautics and Space Administration.

This work has made use of data from the European Space Agency (ESA) mission
{\it Gaia} (\url{https://www.cosmos.esa.int/gaia}), processed by the {\it Gaia}
Data Processing and Analysis Consortium (DPAC,
\url{https://www.cosmos.esa.int/web/gaia/dpac/consortium}). Funding for the DPAC
has been provided by national institutions, in particular the institutions
participating in the {\it Gaia} Multilateral Agreement.

These observations were obtained as visiting astronomers at the Infrared Telescope Facility, which is operated by the University of Hawaii under contract 80HQTR19D0030 with the National Aeronautics and Space Administration.

\clearpage

\bibliographystyle{apj}
\bibliography{Survey_bib.bib}

\clearpage

\begin{table}[]
\caption{Scenarios affecting Spectroastrometric Signals of Ro-vibrational CO Emission Lines}
\label{tab:scenarios}
    \centering
 \begin{tabular}{|l|l|l|l| }
 \hline
Scenario             & Line Profile & Geometry Resulting in SA Signal & Variability Timescale \\
\hline
Circumplanetary Disk & Asymmetric & \multirow{2}{7cm}{A planet plus a circumplanetary disk (CPD) system offset from the star. The CPD emission would be nearly a point source.} & Orbital timescale. \\
& & & \\
& & & \\
\hline
Eccentric Annulus &  Asymmetric & Brighter side of profile more extended. & \multirow{2}{3.5cm}{Persistent for 100's of orbital timescales.} \\
& & & \\
\hline
Vortex &  Asymmetric & \multirow{2}{7cm}{An azimuthally asymmetric circumstellar disk.} & Orbital timescale.  \\
 & & & \\
\hline
Disk Wind & \multirow{2}{3.5cm}{Centrally peaked; Preferentially blueshifted} & Extended along semi-minor axis toward observer. &  \multirow{2}{3.5cm}{Stochastic variability timescale.}  \\
 & & & \\
\hline
 \end{tabular}
\end{table}

\begin{table}[]
\caption{Adopted Parameters}
\label{tab:literature_params}
    \centering
    \begin{tabular}{l c c c c c}
    Star ID & M$_{star}$ & Disk & Semi-Major & Radial & Distance\\
     & & Inclination & Axis PA & Velocity & \\
     & (M$_{\odot}$) & ($^{\circ}$) & ($^{\circ}$) & (km s$^{-1}$) & (pc) \\
    \hline
HD~141569        & 2.00 & 51 & 356 & -6.0 & 110.6 \\
HD~179218        & 3.00 & 40 &  23 & 15.1 & 260 \\
HD~169142        & 1.85 & 13 & 5   & -3.0 & 114 \\
LkH$\alpha$~330  & 2.00 & 28 & 49  & 16.3 & 308 \\
EM*~SR~21A       & 1.66 & 16 & 18  & -7.1 & 178 \\
Oph-IRS~48       & 2.00 & 42 & 95  & -5.7 & 136 \\
CQ~Tau           & 1.57 & 35 & 55  & 15.8 & 162 \\
V892~Tau         & 6.00 & 55 & 52  & 17.2 & 135 \\
AB~Aur           & 2.20 & 20 & 50  & 15.3 & 156 \\
\hline

    \end{tabular}
\end{table}

\begin{table}
\begin{center}
\caption{Observation Details}
\label{tab:obs_param}
\begin{tabular}{l c c c c c c}
Star ID & Date & Slit PA & Slit width & Seeing & Integration time & Resolution\\
& & Parallel/Antiparallel & & &\\
 & & ($\degr$) &  ($^{\prime \prime}$) &  ($^{\prime \prime}$) & (s) & (km s$^{-1}$)\\
\hline
HD~141569       & Apr 19, 2017 & 356 / 176 & 0.375  & 0.75 & 11388 & 6\\
HD~179218       & Jul 7, 2020  & 23, 113 / 203, 283 & 0.375 & 0.55 & 3240  & 3.5\\
HD~169142       & Apr 20, 2017 & 5 / 185  & 0.75  & 0.78 & 2372  & 6\\  
                & Aug 25, 2017 & 5 / 185  & 0.375 & 0.72 & 2135  & 3.5\\
LkH$\alpha$~330 & Aug 25, 2017 & 165 / 345  & 0.75  & 0.81 & 6405  & 6\\
EM*~SR~21A      & Aug 2, 2017  & 16 / 96  & 0.375 & 0.75 & 2372 & 3.5\\
Oph-IRS~48      & Aug 2, 2017  & 95 / 275 & 0.375 & 0.79 & 712 & 3.5\\
CQ~Tau          & Jan 16, 2021 & 47 / 227  & 0.375 & 0.82 & 4448  & 3.5\\
                & Jan 22, 2022 & 137 / 317 & 0.75  & 0.92 & 2700  & 6\\
                & Jan 21, 2023 & 47 / 227 & 0.75  & 0.72 & 2640  & 6\\
V892~Tau        & Jan 16, 2021 & 235 / 55 & 0.375 & 1.23 & 2224  & 3.5\\
                & Jan 21, 2022 & 270 / 90 & 0.75  & 2.06 & 2040  & 6\\
AB~Aur          & Feb 6, 2021  & 60 / 240 & 0.375 & 0.62 & 6240  & 3.5\\
                & Feb 5, 2021  & 24 / 204  & 0.375 & 0.91 & 6960  & 3.5\\
                & Feb 4, 2021  & 150 / 330 & 0.375 & 1.76 & 6960 & 3.5\\
\hline

\end{tabular}

\end{center}
\end{table}

\begin{table}[]
\caption{Model Fit Parameters}
\label{tab:model_fit_params}
    \centering
    \begin{tabular}{l c c c c c c c c c c}
        Star ID & a$_{in}$  & a$_{out}$ & $\alpha_I$ & e$_0$  & $\alpha_e$ & $\omega$   \\
                & (au)      & (au)      & \empty     & \empty & \empty     &  (rad) \\
    \hline
    HD~141569             & 12.5$\pm$0.9 & 43$\pm$2 & -0.11$\pm$0.14 & 0.23$\pm$0.03 & -5.3$\pm$0.6 & 0.26$\pm$0.04 \\
    HD~179218             & 8$\pm$1  & 55$\pm$5 & -1.8$\pm$0.1 & 0.12$\pm$0.06 & -0.8$\pm$0.2 & 0.4$\pm$0.3 \\
    HD~169142             & 2.7$^{+0.6}_{-0.3}$ & 80$\pm$20 & 2.0$\pm$0.1  & $<$0.25 & -3 & 0 \\
    LkH$\alpha$~330       & 0.46$\pm$0.03 & 120$\pm$20 & 2.23$\pm$0.03 & 0.08$\pm$0.03 & -2.3$\pm$0.7 & 3.1$\pm$0.33 \\
    EM*~SR~21A            & 7.3$^{+0.5}_{-0.7}$ & 9.0$^{+2.0}_{-0.5}  $ & 1.4$^{+0.2}_{-1.1}$ & $<$0.13 & -3 & 0   \\
    Oph-IRS~48            & 25$^{+3}_{-5}$      & 41$\pm10$ & 2.0$\pm$1.5 & $<$0.17 & -3 & 0    \\
    CQ~Tau                & 0.51$\pm$0.02 & 9.8$\pm0.7$ & 2.52$\pm$0.03 & 0.27$\pm$0.02 & -5.0$\pm$0.5  & 2.54$\pm$0.05 \\
    V892~Tau$^{\dagger}$  &  7 & 50 & 2.3 & 0.1 & -3 & 0 \\
    AB~Aur                & 1.0$\pm$0.2 & 106$\pm$15 & 2.2$\pm$0.1 & $<$0.07 & -3 & 0  \\
\hline
    \end{tabular}
    
$^{\dagger}$No good fit could be found. Parameters used for example model presented in Figure~\ref{fig:linegallery}.
    
\end{table}

\end{document}